# Validation of daylighting model in CODYRUN building simulation code

H. Boyer[1,a], B. Malet-Damour[1], A.H. Fakra[1], S. Guichard[1], A. Jean[1],
T. Libelle[1], D. Bigot[1], F. Miranville[1], M. Bojić[2,b]

[1] Physics and Mathematical Engineering Laboratory for Energy and Environment (PIMENT), University of La Réunion, 117 rue du Général Ailleret, 97430 Le Tampon, France

[2] Faculty of Engineering, University of Kragujevac, 34000 Kragujevac, Serbia

[a] harry.boyer@univ-reunion.fr, [b] bojic@kg.ac.rs

**Abstract.** CODYRUN is a multi-zone software integrating thermal building simulation, airflow, and pollutant transfer. A first question thus arose as to the integration of indoor lighting conditions into the simulation, leading to a new model calculating natural and artificial lighting. The results of this new daylighting module were then compared with results of other simulation codes and experimental cases both in artificial and natural environments. Excellent agreements were obtained, such as the values for luminous efficiencies in a tropical and humid climate. In this paper, a comparison of the model output with detailed measures is presented using a dedicated test cell in Reunion Island (French overseas territory in the Indian Ocean), thus confirming the interest for thermal and daylighting designs in low-energy buildings.

**Introduction**

Several software packages are available for thermal and airflow simulation in buildings. The most frequently used are ENERGY+ [1], ESP-r [2], and TRNSYS [3]. These applications allow an increasing number of models to be integrated, such as airflow, pollutant transport, and daylighting. In the latter category, we may note ENERGY+, ESP-r and ECOTECT [4] software. After more than 20 years of developing a specific code named CODYRUN, we decided to add a lighting module to our software. This paper therefore provides some details on this evolution and elements of validation.

**The CODYRUN initial software and its validation**

Developed by the Physics and Mathematical Engineering Laboratory for Energy and Environment at the University of Reunion Island, CODYRUN [5-14] is a multi-zone software program integrating ventilation and moisture transport transfer in buildings. The software employs a zone approach based on nodal analysis and resolves a coupled system describing thermal and airflow phenomena. Numerous validation tests of the CODYRUN code were successfully applied to the software. Apart from the daylighting model, the majority applied the BESTEST procedure [15]. The International Energy Agency (IEA) sponsors a number of programs to improve the use and associated technologies of energy. The National Renewable Energy Laboratory (NREL) developed BESTEST, which is a method based on comparative testing of building simulation programs, on the IEA's behalf. The procedure consists of a series of test cases buildings that are designed to isolate individual aspects of building energy and test the extremes of a program. As the modelling approach is very different between codes, the test cases are specified so that input equivalency can be defined thus allowing the different cases to be modelled by most of codes. The basis for comparison is a range of results from a number of programs considered to be a state-of-art in United States and Europe. Associated with other specific comparisons, a very confident level of validation was obtained for the CODYRUN initial software [8].

# Integration and validation of a daylight model

The model and its integration in CODYRUN :

All available daylight models require relative spatial positioning of the elements (floors, windows, openings, and light sources) to calculate the received light on each point in the working plane (or mesh). The data structures initially used by CODYRUN had to be completed with vertexes [15] A set of related C procedures relative to the geometry of planar polygons was also incorporated into the code [2,16,17]. It relates to, for example, the calculation of surface perimeters for polygons, the distance between a point and a plane, testing the inclusion of a point in a plane or polygon, or projections relating to shading calculations and the sunspot. Then, the daylight factor (DF) classical method was used to calculate the diffuse illuminance for all of the mesh points. This method was elaborated by the British Building Research Establishment (BRE) and published by the Chartered Institution of Building Services Engineers (CIBSE) [18]. The diffuse light falling on one point in a room is commonly considered to be composed of three distinct parts: the sky component (SC), externally reflected component (ERC), and internally reflected component (IRC). These values were calculated for each point of the defined mesh using standard formulas. The direct illumination was obtained by projecting the edges of the glazing and the outer apertures eventually illuminated by direct solar radiation in order to calculate the indoor sunspot.

Validation :

When dealing with the daylighting simulation model in particular, it is necessary to refer to some rigorous and recognized procedures used around the world. However, there is limited documentation on the procedures to follow. In most cases, laboratories implement their own experimental database to serve as a reference for comparisons between model predictions and measurements. Covering both artificial and natural lighting, we use Maamari's test cases [19] for the validation step. Indeed, this detailed study establishes methods to verify the reliability of simulation codes for indoor daylighting on the basis of analytical and experimental tests cases. This work was used as a reference in task TC3-33 of the International Commission on Illumination (CIE). Many other test cases found in scientific literature (publications from the French Scientific and Technical Center for Building, BRE, International Energy Agency's Task 21, and experimental test case from CIBSE, etc.) were applied to the CODYRUN simulation software. An inter-software comparison was also conducted, and finally, database references for the local study in dynamic conditions (from a cell called LGI) were established [11] and are briefly presented below.

It is common in the field of lighting to consider at a given time of day the spatial distribution of the illumination, so as to determine whether the required level is sufficient. Other studies are related to longer periods, such as energy consumption or daylight autonomy. In this article, our approach is different and that monitoring with short timestep (hour or minute) lighting at certain points.

An experimental cell 1:1 was instrumented to verify the reliability of the simulation software CODYRUN photometric interior environments. The experimental cell located in Saint-Pierre (Réunion Island, France, 21°19'S, 55°28'E) was initially built to experimentally validate physical models introduced in CODYRUN. The weather station, positioned on the experimental platform, measures solar radiation (global and diffuse) and the wind speed and direction (respectively at a height of 2 m and 10 m above the ground). The datalogger is Campbell-type. Initial measurements were completed by an outdoor luxmeter measuring the global outdoor illuminance values. It is positioned on the roof of the experimental cell LGI. For this study, the instrumentation at the station consists of two pyranometers and illuminance sensors for horizontal solar irradiance and daylight illuminance data measurements.

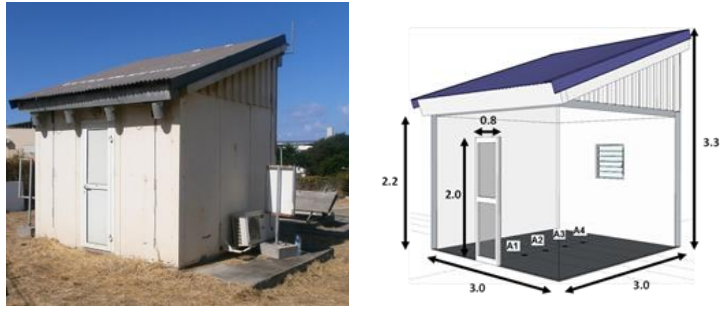

Fig. 1. Picture and sketch of the test cell.

Experiments done on the LGI cell allowed the comparison of simulated daylight values with those measured experimentally. Measurements of internal illuminance were made four points (A1 to A4), 0.01 m above the floor. The sensors were aligned to the central axis of the building (cutting the glazing into two symmetrical parts) and perpendicular to single glazing plane (glazing pane shown obstructed on the left part of previous figure was unobstructed when indoor measurements were made). They are placed in alignment with the openings on the slab as shown also in Figure 1. These sensors are equidistant from 0.5 m. The horizontal distance between the first sensor (A1) position and the opening is 0.73 m. The measurements presented in Figures 2 (global, direct and diffuse radiation) and 3 (indoor illuminance) were taken during the day of 10 February 2008, with overcast sky. Longer sequences and discussion are available in [11].

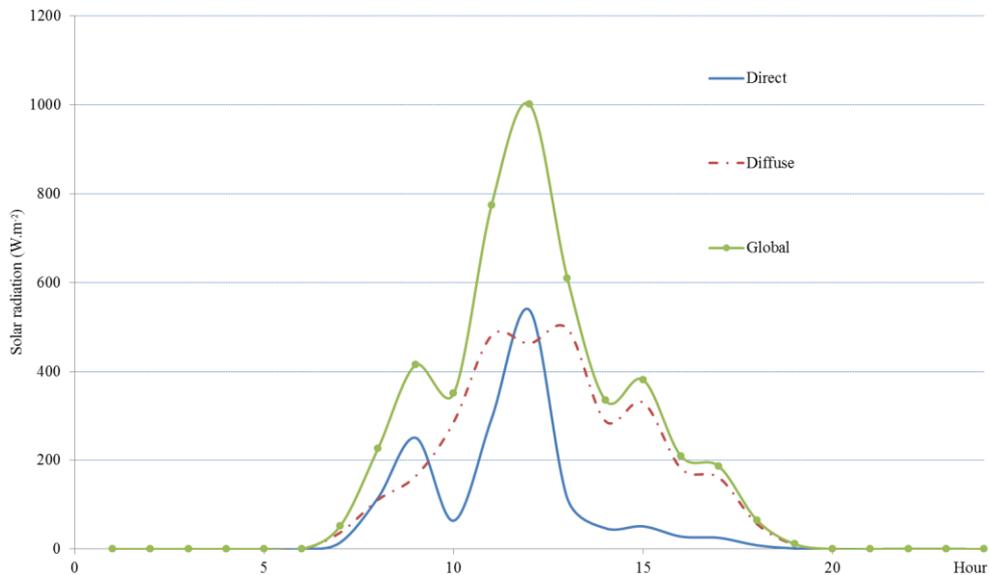

Fig.2. Outdoor radiation measurements.

It is very common in the field of lighting to consider for a given hour the spatial distribution of the illumination, so as to determine whether the required level is sufficient. Other studies are related to longer periods, such as energy consumption or daylight autonomy, as [14]. In this article, our approach is different and that monitoring with short timestep (hour or minute) lighting at certain points. On next figure, values encountered at A1 location are much greater than at A3, due to shorter distance from glazed door. In most cases, simulated values in CODYRUN (labelled CODYRUN_A1 and CODYRUN_A3) between 10:30 am and 12:45 pm are higher compared to measured values (Measures_A1 and Measures_A3). The explanation is that the studied cell is equipped with an overhang on the northern façade. This prevents a major part of the sunlight from penetrating into the room. At this time, this particular device is not taken into account by the software. Concerning the evening, we noticed that the simulated values are lower than the measured values. The fact is that South and West parts of vertical façades of the test building are opaque.

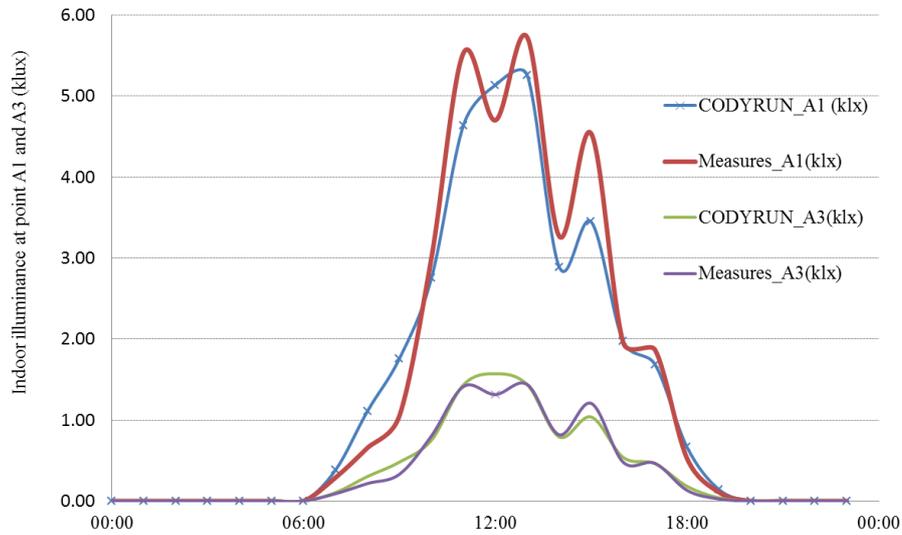

Fig.3. Indoor illuminance simulation and measurement at A1 and A3 points.

Then, the North-facing overhang does not influence skylight at these hours of the day. We obtain more coherent values between the simulation and the measurements and model is clearly sensitive to parameters as the ground albedo. The results show, on the whole day, a good accuracy of the software in the calculation of the global indoor illumination, except for the most distant points from the opening.

**Conclusion**

Our goal of integrating a model for lighting calculation was reached, with some details on the internal organization provided in this article. The daylight studies can be performed using other software, but as in previous CODYRUN embedded models (pollutants, airflow, etc.), we believe our own coding for this lighting model to be profitable in terms of its in-depth knowledge and flexibility, as it takes into account certain specificities or local practices. The results of actual release are satisfactory in terms of their precision and comparability to much more powerful software. They allow us, in the philosophy of CODYRUN, to conduct specific software development dedicated to research and professional audiences. In future research, we aim to explore the innovative integration of specific components for lighting (lightpipes, thermotropic layers, etc.) and interfacing the software with Google Sketchup® and OpenStudio® to improve its usability among professionals.